\pdfoutput=1

\documentclass[11pt]{article}
\usepackage{geometry}
\usepackage{tikzit}

\usepackage[affil-it]{authblk}

\usepackage[hyphens]{url}
\usepackage{hyperref}
\usepackage[hyphenbreaks]{breakurl}

\hypersetup{
    colorlinks,
    linkcolor={red!80!black},
    citecolor={red!80!black},
    urlcolor={red!80!black}
}

\usepackage[
    sorting=none,
    backend=biber,
    maxbibnames=4,
    doi=true,
    isbn=false,
    url=true,
]{biblatex}

\DeclareSourcemap{
  \maps[datatype=bibtex]{
    \map{
      \step[fieldsource=doi,final]
      \step[fieldset=url,null]
      \step[fieldset=urldate,null]
    }
    \map{
      \step[fieldsource=eprint,final]
      \step[fieldset=url,null]
      \step[fieldset=urldate,null]
    }
  }
}

\usepackage[frozencache,cachedir=./minted-ms]{minted}
\newcommand{\py}[1]{{\color{brown}\mintinline{python}{#1}}}
\setminted{frame=lines}

\BeforeBeginEnvironment{minted}{\vspace{-16pt}}
\AfterEndEnvironment{minted}{\vspace{-8pt}}

\usepackage{amsthm}

\newtheorem*{example}{Example}

\title{DisCoPy for the quantum computer scientist}

\author{Alexis Toumi$^{\star \dagger}$}
\author{Giovanni de Felice$^\dagger$}
\author{Richie Yeung$^\dagger$}

\affil{
$\dagger$ Cambridge Quantum Computing Ltd. \\
$\star$ Department of Computer Science, University of Oxford
}

\begin{document}

\maketitle

\begin{abstract}
DisCoPy (Distributional Compositional Python) is an open source toolbox for computing with string diagrams and functors.
In particular, the diagram data structure allows to encode various kinds of quantum processes, with functors for classical simulation and optimisation, as well as compilation and evaluation on quantum hardware.
This includes the ZX calculus and its many variants, the parameterised circuits used in quantum machine learning, but also linear optical quantum computing.
We review the recent developments of the library in this direction, making DisCoPy a toolbox for the quantum computer scientist.
\end{abstract}

String diagrams first appeared in the work of Hotz~\cite{Hotz65} where they describe how digital circuits compose in sequence and in parallel.
Penrose~\cite{Penrose71} then reinvented them as a graphical alternative to the abstract index notation for tensors and spinors, which he then used in his work on general relativity with Rindler~\cite{PenroseRindler84}.
More recently, they played a prominent role in categorical quantum mechanics~\cite{AbramskyCoecke04} and the ZX-calculus~\cite{CoeckeDuncan11} where they provide a formal graphical language for composing quantum processes in sequence and in parallel.
From a piece of notation used in mathematical papers, string diagrams have become a data structure used in software applications.
On the one hand, PyZX~\cite{KissingerVanDeWetering19} performs quantum circuit optimisation via automated rewriting of ZX-diagrams.
It is a high-performance computing tool, but it is tailored for a specific application and it cannot handle arbitrary diagrams.
On the other hand, the graphical proof assistant homotopy.io~\cite{ReutterVicary19} can represent arbitrary $n$-dimensional diagrams, but it does not perform any computation other than its user interface.

DisCoPy~\cite{FeliceEtAl20a} (Distributional Compositional Python) is an open source library that aims to get the best of both worlds: it is general enough to encode arbitrary string diagrams and it also allows to interpret these diagrams as computation.
It comes with an extensive documentation\footnote{\url{https://discopy.readthedocs.io/}} and many interactive examples in the form of Jupyter~\cite{KluyverEtAl16} notebooks.
DisCoPy diagrams can be evaluated as high-performance classical code via interfaces with
NumPy~\cite{VanDerWaltEtAl11},
JAX~\cite{BradburyEtAl20},
TensorFlow~\cite{AbadiEtAl16},
PyTorch~\cite{PaszkeEtAl19}
and TensorNetwork~\cite{RobertsEtAl19} as well as symbolic computation via Sympy~\cite{MeurerEtAl17}.
The same diagram data structure can be evaluated on quantum hardware via the t$\vert$ket$\rangle$ compiler~\cite{SivarajahEtAl20}, it is also compatible with the optimisation algorithms of PyZX~\cite{KissingerVanDeWetering19}.
While the library is a general-purpose toolbox, its development was driven by its use in quantum natural language processing (QNLP)~\cite{CoeckeEtAl20,MeichanetzidisEtAl20a}, an application which was later packaged into its own specialised library, lambeq~\cite{KartsaklisEtAl21}.
This report gives a short presentation of DisCoPy as a toolbox for the quantum computer scientist and a review of its recent developments.


We refer the reader interested in formal definitions to the article introducing DisCoPy~\cite{FeliceEtAl20a}.
Rather than defining what diagrams are, we will describe what the user can do with them.
Their implementation is given by \py{Diagram}, a Python class with the following interface:
\begin{itemize}
\item \py{Diagram.dom} and \py{Diagram.cod} are attributes of type \py{Ty}, a class which stores the list of labels for the input and output wires, also called the domain and codomain,
\item \py{Diagram.id}, shortened to \py{Id}, is a static method which takes an \py{x: Ty} as input and returns the identity diagram with \py{dom = cod = x}, i.e. \py{x}-labeled wires in parallel,
\item \py{Diagram.then} and \py{Diagram.tensor}, shortened to \py{>>} and \py{@}, are methods which take two diagrams and return their composition in sequence and in parallel respectively,
\item \py{Diagram.dagger}, shortened to \py{[::-1]}, returns the vertical reflection of a diagram,
\item \py{Diagram.draw} is a method which either displays the diagram on screen using matplotlib~\cite{Hunter07} or outputs TikZ~\cite{Tantau13} code to be included in \LaTeX \ documents,
\item \py{Box} is a subclass of \py{Diagram} defined by three attributes \py{name: str}, \py{dom: Ty} and \py{cod: Ty}.
\end{itemize}
The \py{Diagram} class is general enough to encode any string diagram, and in particular any quantum circuit.
Indeed, its subclass \py{Circuit} comes with methods \py{from_tk} and \py{to_tk} for back and forth conversion into the circuit data structure of the t$\vert$ket$\rangle$ compiler~\cite{SivarajahEtAl20}.
Pure circuits have domain and codomain given by powers \py{qubit ** n} of a single object \py{qubit: Ty}.
\py{Gate}, \py{Bra} and \py{Ket} are subclasses of \py{Box} and \py{Circuit} that encode unitary quantum gates, post-selected measurement and state preparation respectively.

Apart from drawing them, we cannot do much with diagrams alone.
We need to call a \py{Functor} to interpret them in terms of (classical or quantum) computation.
This is defined by two attributes \py{ob} and \py{ar} for the interpretation of wires and boxes respectively.
In particular, a \py{tensor.Functor} interprets wires as dimensions and boxes as arrays so that calling the functor on a diagram performs tensor contraction.
Thus, DisCoPy implements classical simulation of pure quantum circuits in just a few lines of Python, wrapped into the \py{Circuit.eval} method which calls NumPy by default or any chosen high-performance backend.

\begin{example}
Let's simulate a post-selected quantum teleportation protocol.
{\normalfont
\begin{minted}{python}
from discopy.quantum import qubit, Id, Bra, Ket, CX, H, scalar
Bell_state = Ket(0, 0) >> H @ Id(qubit) >> CX
protocol = Ket(1) @ Bell_state >> Bell_state.dagger() @ Id(qubit)
result = Ket(1) @ scalar(.5)  # The protocol has .5 probability of success.
assert protocol.eval() == result.eval()
\end{minted}
}
\end{example}


A major feature that was released after the first DisCoPy article~\cite{FeliceEtAl20a} is the implementation of mixed classical-quantum circuits.
This updated \py{Circuit} class comes with wires of type \py{bit: Ty}\footnote
{DisCoPy implements the classical simulation of qudits and digits of arbitrary dimension, with qubit and bit as special cases.
As of today, only qubit circuits can be compiled onto quantum devices.} and boxes \py{Measure} and \py{Encode}\footnote
{The current interface with t$\vert$ket$\rangle$ handles \py{Measure} but not \py{Encode}, which allows classical control of a gate depending on the result of a measurement.
This is required to implement routines such as quantum assertions~\cite{LiEtAl20}.} going from \py{qubit} to \py{bit} and back.
The \py{Discard} box allows to trace out a qubit or to marginalise out a bit.
The method \py{Circuit.eval} computes a classical simulatation of these classical-quantum circuits via Selinger's CPM construction~\cite{Selinger07}.
The result is a \py{CQMap}, the class that implements classical-quantum maps (also called channels) as defined by Coecke and Kissinger~\cite{CoeckeKissinger17}.
Again, this simulation can be computed using any high-performance backend for tensor contraction.
The same \py{Circuit.eval} method can also compile and execute circuits on a quantum device via t$\vert$ket$\rangle$.
Thus, a one-line change can turn a classical simulation into a quantum experiment.
Most of the usual quantum gates are implemented, as well as standard ansätze such as instantaneous quantum polynomials (IQP)~\cite{ShepherdBremner09}.

The \py{Circuit} class can also encode any parameterised quantum circuit, simply by using SymPy expressions as parameters for quantum gates.
The method \py{Circuit.grad} implements diagrammatic differentiation, a method introduced by the authors in a previous article~\cite{ToumiEtAl21} which computes the gradients of any parameterised circuit as a formal sum of circuits.
Diagrammatic differentiation applies to quantum circuits but also to their classical post-processing e.g. by a neural network.
Both can be encoded in the same diagram using the \py{Bubble}\footnote
{Bubbles are to diagrams what brackets are to formulae, they denote a function applied to a diagram inside.} class, the gradient of which is computed by a diagrammatic chain rule.
Thus, DisCoPy can be used as a high-level language for variational quantum algorithms and quantum machine learning.

Diagrammatic differentiation looks even more beautiful when applied to the ZX-calculus, especially in its algebraic variant as introduced by Wang~\cite{Wang20}.
Due to the topology of the algebraic ZX-calculus\footnote
{Because of its asymmetric triangle nodes, the algebraic ZX-calculus lacks a property called flexsymmetry~\cite{Carette21}.},
it cannot be encoded in the graph-like data structure of PyZX.
Using DisCoPy however, it is straightforward however to implement as a subclass of \py{Diagram} with \py{Triangle} boxes as well as \py{Z} and \py{X} boxes for spiders.
The same is true for many other variants: the ZW-calculus used to prove completeness for ZX~\cite{NgWang17} and to encode fermionic quantum circuits~\cite{FeliceEtAl19a}, the ZH-calculus~\cite{BackensKissinger19} for classical non-linearity, scalable ZX~\cite{CaretteEtAl19} to describe registers of qubits, etc.
In a nutshell, DisCoPy can encode the diagrams of any graphical calculus and compute their interpretation as computational processes.
For now, we have implemented the standard ZX-calculus as a subclass \py{zx.Diagram} with methods \py{to_pyzx} and \py{from_pyzx} for automated rewriting with PyZX, and \py{circuit2zx} for translating a \py{Circuit} into a ZX diagram.

An exciting direction of work in progress is to go away from the world of qubits altogether and implement the diagrams for linear optical quantum computing.
DisCoPy's \py{photonics} module (still under development) will allow to build linear optical circuits and interface with Perceval~\cite{HeurtelEtAl22} for their high-performance classical simulation.
Implementing recent theoretical progress on a graphical calculus for linear optics~\cite{FeliceCoecke22}, DisCoPy functors will allow to compile qubit circuits and ZX diagrams into linear optical circuits, one step on the photonic roadmap towards fault-tolerant quantum computing.

\paragraph{Acknowledgements.}
AT thanks Simon Harrison for the Wolfson Harrison Quantum Foundation Scholarship.
We thank everyone who contributed to DisCoPy, even for a few lines: Bob Coecke (who capitalised the name DisCoPy), Ian Fan, Nicola Mariella, Konstantinos Meichanetzidis, Irene Rizzo, Alex Koziell-Pipe, Charlie London and Thomas Hoffmann.

We look forward to welcoming new contributors!

\printbibliography
\end{document}